# Ultrafast light-induced softening of chalcogenide thin films above the rigidity percolation transition

Pritam Khan[1,2], Rajesh Kumar Yadav[1] and K.V. Adarsh[1]*

[1]Department of Physics, Indian Institute of Science Education and Research, Bhopal 462066, India.

[2]Department of Physics, Kyushu University, Fukuoka 819-0395, Japan

*Little is known about the role of network rigidity in light-induced structural rearrangements in network glasses due to a lack of supporting experiments and theories. In this article, we demonstrate for the first time the ultrafast structural rearrangements manifested as induced absorption (IA) over a broad spectral range in a-$Ge_xAs_{35-x}Se_{65}$ thin films above the mean-field rigidity percolation transition, quantified by the mean coordination number ⟨r⟩ = 2.40. The IA spectrum arising from self-trapped excitons, induced structural rearrangements by softening the glass network that strikingly reveal two relaxation mechanisms which differ by one order of magnitude. The fast kinetics of electron-lattice interaction occurs within 1 ps, exhibits a weak dependence on rigidity and dominates in the sub-bandgap region. In a stark contrast, the slow kinetics are associated with the structural changes in the bandgap region and depends strongly on network rigidity. Our results further demonstrate that amplitude of IA scales a linear relationship with excitation fluence which provides a unique way to induce structural rearrangements in over-coordinated network to exploit it for practical purposes. Our results modify the conventional concept of rigidity dependence of light-induced effects in network glasses, when excited with an ultrafast laser.*

---

* Author to whom correspondence should be addressed; electronic mail: adarsh@iiserb.ac.in

# Introduction

The interplay between light-induced effects and network rigidity is a heavily debated topic in several areas of contemporary physics, including optics, photonics, and glass science [1-3]. Therefore, it is quite alluring to study the influence of topological constraints on physical properties as a function of network rigidity. The results of this fundamental research may have a tremendous impact on the design of novel optoelectronic-based technologies [4-6]. Chalcogenide glasses (ChGs) are canonical systems that are used to study the influence of topological constraints on physical properties as a function of network rigidity [7].

Se-As-Ge system is of particular interest because the covalent coordination numbers varies from 2 to 4 along the series by changing the local structure from polymer chain-like selenium to three-dimensionally rigid network glasses [8]. Plus, they form a close to ideal network because of the similar size and electronegativity of the elements [9]. Apart from that, Ge-As-Se glasses allow composition variation over wide range and therefore studied by several research groups over the past few decades [10-13]. According to the network rigidity theory by Phillips and Thorpe [14, 15], mean coordination number (MCN), characterized by $\langle r \rangle$, plays an important role in determining the extent of lightinduced effects of a covalent network. The theory predicts that, at $\langle r \rangle$=2.40, the number of degrees of freedom of a network exactly equals the number of constraints and the network makes a rigidity percolation transition from under-constrained "floppy" to over-constrained "rigid" network. This was further supported with mean field theory [14, 15] and computer simulations.

In our present study, we have chosen $Ge_xAs_{35-x}Se_{65}$ glasses, for which MCN ($\langle r \rangle$) is given by

$$\langle r \rangle = 4x + 3(35-x) + 2 \times 65 \tag{1}$$

Two general themes have evolved to explain the light-induced effects: the first one emphasizes the aspects of the local structure and the second one stresses on the global structure [16]. In the former theme, bandgap light illumination produces electron-hole states known as excitons that get self-trapped in the deformable glassy network. The self-trapping of excitons can be attributed to the formation of metastable local defect configurations of under- and over-coordinated chalcogen atoms in the neutral and charged states [17-20]. These local structural changes form localized states in the tail of the band edges and results in a reduction of the bandgap. With prolonged illumination, these defects mediate rapid switching of normal covalent bonds leading to photo-diffusion, which in turn results in global structural changes [21]. Global and local structural changes are associated with the softening of the amorphous matrix and therefore connected to the network rigidity characterized by ⟨r⟩. Therefore, it is expected that the light-induced effects vanish above the mean-field rigidity percolation transition of ⟨r⟩ = 2.40. Based on thermodynamic measurements, it has been shown that the thermal and mechanical properties of Ge-As-Se network glasses are modified significantly with ⟨r⟩. For example, thermal properties such as specific heat, enthalpy [22] and thermal relaxation [9] show a global minimum at ⟨r⟩ = 2.40. Likewise, light-induced effects such as photodarkening, photoexpansion, and photorelaxation vanish at ⟨r⟩ = 2.40 [9], a parallel behavior to the thermal properties. Such results imply that perhaps the interaction of light with a glassy network may intrinsically depend on its rigidity. However, mechanical relaxations which are depicted by barrier energy along with viscosity, exhibit a "U" shaped behavior near ⟨r⟩ = 2.40 [23]. Strikingly, in our recent work [24], we observed that photobleaching, which is an opposite phenomenon to photodarkening, is maximum for a rigid system and in particular at ⟨r⟩ = 2.60. However, later we confirmed by using structure-specific FIR and Raman spectroscopy that this behavior is due to surface photo-oxidation of Ge atoms and not because of structural rearrangements.

In a stark contrast to previous theoretical and experimental results, in this article, we show for the first time that ultrafast structural rearrangements in the over-constrained a-$Ge_xAs_{35-x}Se_{65}$ films when MCN is varied from 2.40 to 2.60, i.e., MCN is above the mean-field rigidity percolation transition. Our results further demonstrate that the structural changes are strongly correlated to the ⟨r⟩ and modify the conventional concept on the dependence of network rigidity on light-induced effects. Kinetic analysis using global target analysis reveals two distinct relaxation mechanisms. The first one is weakly rigidity dependent fast kinetics which occurs approximately within 1 ps and is attributed to electron-lattice interaction. The second slow relaxation process scales with network rigidity and is associated with structural rearrangements due to the self-trapping of excitons [2, 17, 25].

## Materials and Methods

To study the ultrafast light-induced structural rearrangements in the over-constrained network, we have prepared a-$Ge_xAs_{35-x}Se_{65}$ (where x = 5, 10, 15, and 25) thin films of thickness ∼700 nm on a microscope glass substrate by conventional thermal evaporation in a vacuum of $5 \times 10^{-6}$ mbar. We maintained a deposition rate of 2-5 Å/s to preserve the target stoichiometry between the bulk glasses and the thin films which was later confirmed with the Energy-dispersive X-ray. Each sample was associated with their respective MCN as follows: Sample#1: $Ge_5As_{30}Se_{65}$: MCN = 2.40, Sample#2: $Ge_{10}As_{25}Se_{65}$: MCN = 2.45, Sample#3: $Ge_{15}As_{20}Se_{65}$: MCN = 2.50 and Sample#4: $Ge_{25}As_{10}Se_{65}$: MCN = 2.60. Clearly, the MCN of all the samples lies above the mean-field rigidity percolation threshold of ⟨r⟩ = 2.40.

Before studying ultrafast induced absorption (IA), we initially recorded the ground state optical absorption (OA) spectrum of the samples as shown in Fig. 1(a). It is quite evident from the figure that OA spectrum is blue shifted when the MCN increases from 2.40 to 2.60. Next, we calculated the bandgap ($E_g$) of the samples using classical Tauc plots and found that $E_g$ is 1.82 ± 0.01, 1.88 ± 0.01, 1.95 ± 0.01 and 2.07 ± 0.01 eV for MCN 2.4, 2.45, 2.5 and 2.6 respectively. For the Tauc calculation, we have selected the wavelength region where the absorption coefficient (α) is more than $10^4$ cm$^{-1}$. By choosing the α (α = ΔA/d, where ΔA is the absorbance) we can disregard the effects that can arise from different sample thickness.

To study ultrafast IA, we employed femtosecond pump-probe transient absorption spectroscopy. The fundamental beam of 120 fs pulses centered at 800 nm with a repetition rate of 1 kHz, was split into two beams to generate pump and probe. The first beam was passed through an optical parametric amplifier to generate the pump beam. The second beam was delayed with a computer-controlled motion controller and then focused into a CaF$_2$ plate to generate a white light continuum (450- 850 nm) which was used as the probe beam. The probe beam was overlapped within the pump beam for uniform illumination. In our experiments, we have used pump beams of 400 and 560 nm, both of which lie above the bandgap region of the samples. Notably, the penetration depth of 400 nm 560 nm laser is not exactly coherent with the sample thickness, so we can assume that the observed light induced effects are mostly originating from the surface. We define IA of the probe beam ΔA= log[$I_{ex}$(p)/$I_0$(p)] − log[$I_{ex}$(r)/$I_0$(r)]. The symbols p and r correspond to the probe and the reference, respectively. $I_{ex}$ and $I_0$ are the transmitted intensities of the sequential probe pulses after a delay time τ following excitation by the pump beam and in the ground state respectively [18, 20]. It is important to note that all the pump-probe measurements are performed in ambient condition. To exclude any heating effects, the sample was rotated continuously. We find that temperature rise at the sample spot is approximately 10 K, so we can exclude the possibility of any thermal effect in our samples.

**Results and Discussion**

At first, we excited the sample with the highest MCN = 2.60 with the above bandgap (400 nm) pump and recorded IA at different probe delays. The data is mapped in the contour plot shown in Fig. 1(b). It is quite evident from the figure that the stiffest sample in our series exhibits significant IA. The spectral features of the IA reveal (1) strong and fast decaying IA in the sub-bandgap region (650-720 nm) and (2) weak and slow decaying IA near the bandgap region (540-630 nm). Such result provides the first experimental evidence of ultrafast light-induced effects in an over-constrained rigid network. To check the consistency of our observation, we illuminate the sample with 560 nm pump and observe similar spectra as shown in Fig. 1(c). The observed broad IA is identified as due to self-trapped excitons, principally on the basis of the decay-times. For instance, pump beam triggers the electrons from the valence to the conduction band. After excitation, electrons in the conduction band interact with each other by exchanging energy which is known as electron-electron ($\tau_{el-el}$) interaction, or electron thermalization [26]. For ChGs, the thermalization time is calculated to be less than 100 fs, which is beyond the scope of our measurement [27]. During the carrier thermalization, the electrons and holes diffuse to a distance R which is much smaller than the critical distance $R_C$, known as the Coulomb capture (Onsager) Radius defined as

$$e^2 / 4\pi\varepsilon\varepsilon_0 R_C = kT_0 \quad (1)$$

For Ge-As-Se network glasses, the dielectric constant $\varepsilon$ is approximately 15, and the glass transition temperature $T_0$ is nearly 520 K [28]. After inserting these parameters into Eqn. [1], we found that $R_C$ is approximately 20 A$^o$. In an earlier analysis, the effective mass approximation model indicated that the separation between the e-h pair is about 3 A$^o$ in the ground state [29]. When excited with a femtosecond laser, the separation is increased by a factor of $\sqrt{\beta\tau}$, where $\beta$ is the diffusion constant and $\tau$ is the thermalization time. For ChGs, $\tau$ is ~ fs and $\beta = 1 \times 10^{-3}$ cm$^2$/sec [30], from which we found that $\sqrt{\beta\tau}$ is < 1 A$^o$. This implies the e-h separation is less than the critical distance $R_C$ even after excitation, which strongly favors e-h pairs to form excitons ($E_x$). Subsequently, exciton recombination can take place via two paths: (I) directly to the ground state or (ii) via a metastable state formed by the defect pairs $C_3^+$, $C_1^-$ (known as valence alternation pair (VAP). Here "*C*" is the chalcogen atom, the super and subscripts denote the charge state and the coordination of the defect states respectively. Such decay processes can be well-understood based on the configurational coordinate diagram [2, 31, 32] shown in Fig. 2. For ChGs, exciton recombination via path (II) is more probable due to its low photoluminescence efficiency. If we assume that the formation energy of the defect pair is $E_c$, the amount of energy released in VAP formation is ($E_x$-$E_c$), which strongly accelerates non-radiative recombination. Consequently, the material is left in a different configuration from the initial state. Specifically, the energy released is used to modify the local structure by softening the network which gives rise to change in the bandgap. Strikingly, we observed that structural rearrangements that induce changes in the bandgap region are metastable and decay slowly. However, the VAP created by the pump beam produce excited state absorption in the sub-bandgap region since optical transitions are allowed from the defects to the conduction and valence bands or among themselves at a much lower energy than the bandgap. IA in the defect region decays much faster because they self-annihilate each other [18]. In a simplistic model, we demonstrate that the cumulative effect of the changes in the local configuration and defect creation are responsible for the broad IA because of two dynamics processes. One occurs at the bandgap and the other in the sub-bandgap region. They are simultaneous and attribute to parallel processes rather than sequential because the components in the bandgap increase when the components in the sub-bandgap increase.

To explore the global picture of ultrafast light-induced structural changes above the mean-field rigidity percolation transition, we systematically measure the IA of the samples with ⟨r⟩ ≥ 2.40. In this context, Fig. 3(a) shows the IA spectrum of all samples at a probe delay of 1 ps when illuminated with 400 nm pump pulses. Apparently, IA is determined to be dominant in the sub-bandgap region decays much faster than the bandgap region for all samples. This was accomplished using the plot shown in Fig. 3(b) of the IA spectrum at a probe delay of 30 ps. At this timescale, the IA spectrum mainly contains the contribution from the bandgap region where its tail has already decayed. Moreover, we observed a decrease in the magnitude of the IA from ⟨r⟩ = 2.40 to ⟨r⟩ = 2.60 which indicates that light-induced effects and resulting softening of the network depend strongly on the extent of self-trapping that relies on ⟨r⟩. At this point, our results are similar to the observation for continuous wave light-induced effects in the samples below the mean-field rigidity percolation transition, as demonstrated by Calvez et al. [9] and Lucas et al. [33]. Nonetheless, their studies revealed that the light-induced effects completely vanish for the samples with ⟨r⟩ > 2.40. Since the under-constrained glasses (⟨r⟩ < 2.40) possess a large fraction of zero-frequency modes which allow substantial structural rearrangements between "*on*" and "*off*" states of light illumination and results in large photo-structural changes.

At this point, to provide a consistent picture of ultrafast structural rearrangements, we performed the pump-probe measurements for 560 nm excitation at different fluences of the pump beam. The results are shown in Fig. 4(a) for the stiffest sample of ⟨r⟩ = 2.6. Clearly, we could observe manifold increase in the amplitude of IA in both bandgap and sub-bandgap region with increasing fluence. To provide a quantitative and overall picture, we have plotted in Fig. 4(b) and (c) the variation of IA amplitude of all samples in the bandgap and sub-bandgap region respectively. Our results indicate that IA scales a linear relationship with excitation fluence which can be used as an effective tool used to lift up the constraints and induce large structural changes in over-coordinated glasses.

After identifying the spectral features of the IA and their possible origin, we now consider the detailed relaxation kinetics of IA. To accomplish our goal, we first performed single value decomposition (SVD) using the equation

$$\Delta A(\nu, t) = \sum_{i=1}^{n} A_i(t) W(\upsilon) \sigma_i \qquad (2)$$

where ΔA (ν, t), $A_i(t)$, W(ν) and $σ_i$ are the IA data, the amplitude as a function of time, the amplitude as a function of frequency and the singular values, respectively. Based on the analysis, we found that two decay constants have significant contributions to the IA data. Then the global analysis schema is performed by fitting the experimental data using a linear combination of 2 exponentials given by:

$$\Delta A(\lambda, t) = i(t) \otimes \sum_{1}^{2} a_l(\lambda) \exp(-t/\tau_l) \qquad (3)$$

Where $i(t)$, $α_l(λ)$, $τ_l$, and $⊗$ are the instrument response function, the amplitude of the exponential decay function, the characteristic time constant to decay to 1/e of its original value and the convolution operator, respectively. The corresponding decay constants from the best fit to the experimental data are $τ_1 = 3.3 ± 0.4$ ps and $τ_2 = 417 ± 3$ ps for MCN = 2.4, $τ_1 = 2.5 ± 0.5$ ps and $τ_2 = 355 ± 3$ ps for MCN = 2.45, $τ_1 = 3.3 ± 0.4$ ps and $τ_2 = 271 ± 3$ ps for MCN = 2.5, and $τ_1 = 1.8 ± 0.2$ ps and $τ_2 = 151 ± 2$ ps for MCN = 2.6. The global analysis procedure emphasizes the common lifetimes for the IA decay because they are parallel in nature. Nonetheless, the amplitude of the associated exponential functions ($α_l(λ)$) of both decay constants will vary depending on the amount of each species that is present. A plot of the amplitude of the exponential function against wavelength, weighted by the lifetime, will produce an intensity-wavelength graph which we defined as the decay associated spectrum (DAS) for a particular lifetime. We have shown in Fig. 3(c) and (d) the DAS spectra that correspond to decay constants $τ_1$ and $τ_2$ respectively. It can be seen from Fig. 3(c) that for all samples, the amplitude of the DAS1 spectra ($τ_1$) is significantly higher in the sub-bandgap region than in the bandgap region. In addition, we also determined from Fig. 3(d) that except for MCN = 2.4, the amplitude of the DAS2 spectra ($τ_2$) in the bandgap region is higher or comparable to that of the sub-bandgap region. This is in stark contrast to the amplitude variation of the DAS1 spectra shown in Fig. 3(c) and clearly indicates that the IA in the bandgap region decays slowly because structural rearrangement takes place. We also found that the amplitude of the DAS2 in the bandgap region is a maximum for MCN = 2.4, which supports our claim that lightly-constrained networks undergo the largest structural rearrangement.

To provide a comprehensive picture of the electron-lattice interaction and the ultrafast structural rearrangements associated with DAS1 and DAS2 respectively, we have selected the central wavelength of the DAS spectra and fitted the experimental data using Eq. [3]. In this context, to get an overall picture, a schematic representation of the timescales of the predominant excitation and relaxation processes in a-$Ge_xAs_{35-x}Se_{65}$ networks for illumination with femtosecond laser pulses, is shown in Fig. 5(a). The temporal evolution of the kinetics curves is shown in Fig. 5(b) and (c) for DAS1 and DAS2, respectively. It can be seen from Fig. 5(b) that the kinetics curves are almost indistinguishable for MCN = 2.5 and 2.6. However, the curves are least steep for MCN = 2.4, that clearly indicates that electron-lattice interactions become slower in lightly-constrained network compared to over-constrained ones. Our result is further supported by the decay time constants ($\tau_1$), that was previously indicated for the different samples. Importantly, we found from Fig. 5(b) that the faster process of electron-lattice interaction represented by DAS1 is fully reversible in nature. However, Fig. 5(c) shows that the slope of the kinetics curves associated with DAS2 is steeper while transitioning from ⟨r⟩ = 2.40 to 2.60. This indicates that exciton recombination and the associated structural changes occur faster for over-constrained networks compared to lightly-constrained ones. Our conclusion is further supported from Fig. 5(d) which reveals that $\tau_2$ decreases linearly with MCN. Notably, in an over-constrained system, the restoring forces are much stronger compared to the lightly-constrained network. As a result, a rigid network can easily return to its initial configuration. Likewise, reversibility of the slower component of IA associated with structural rearrangements, which is represented by DAS2 will exhibit a strong dependence on network rigidity. Consequently, we found that the non-reversible part of the IA is largest for lightly-constrained network of ⟨r⟩ = 2.4 and decreases gradually during the transition to over-constrained network of ⟨r⟩ = 2.6. Such observation clearly suggests that permanent structural change is dominant in lightly-constrained systems compared to over-constrained networks.

## Conclusions

In summary, our results provide the first observation of ultrafast structural rearrangements in an over-constrained rigid a-$Ge_xAs_{35-x}Se_{65}$ network, above the mean rigidity percolation transition at ⟨r⟩ = 2.40. A detailed global analysis of kinetic data shows the existence of two parallel decay mechanisms. The faster process of electron-lattice interaction is dominating in the sub-bandgap region. On the other hand, the slower process, which is manifested by the ultrafast structural rearrangements, i.e. softening of the network, occur near the bandgap of the sample and depend strongly on the network rigidity. The structural rearrangement is further associated with the self-trapping of excitons which is found to maximum for lightly-constrained network. The outcomes of our experimental results can be extended for other chalcogenide systems as well. We believe that, a detailed understanding of the interplay between ultrafast lightinduced effects and network rigidity may pave the way to exploit it for practical purposes specially in the field of glass sciences.

## Acknowledgements

The authors thank Science and Engineering Research Board (Project no: EMR/2016/002520), Council of Scientific and Industrial Research, India, (grant No. 03 (1250)/12/EMR-II) and DAE BRNS (Sanction no: 37(3)/14/26/2016-BRNS/37245) for their financial support.

# References

[1] A. Mishchenko, J. Berashevich, K. Wolf, A. Reznik, D. A. Tenne, and M. Mitkova, Opt. Mater. Express **5**, 295 (2015).

[2] P. Khan, T. Saxena, H. Jain, and K. V. Adarsh, Opt. Lett. **40**, 768 (2015).

[3] X. Feng, W. J. Bresser, and P. Boolchand, Phys. Rev. Lett. **78**, 4422 (1997).

[4] M. Bapna, R. Sharma, A. R. Barik, P. Khan, R. R. Kumar, and K. V. Adarsh, Appl. Phys. Lett. **102**, 213110 (2013).

[5] M. Wuttig, and N. Yamada, Nat. Mater. **6**, 824 (2007).

[6] A. Zakery, Y. Ruan, A. V. Rode, M. Samoc, and B. L. Davis, J. Opt. Soc. Am. B **20**, 1844 (2003).

[7] M. T. M. Shatnawi, C. L. Farrow, P. Chen, P. Boolchand, A. Sartbaeva, M. F. Thorpe, and S. J. L. Billinge, Phys. Rev. B **77**, 094134 (2008).

[8] R. P. Wang, D. Bulla, A. Smith, T. Wang, and B. L. Davies, J. Appl. Phys. **109**, 023517 (2011).

[9] L. Calvez, Z. Yang, and P. Lucas, Phys. Rev. Lett. **101**, 177402 (2008).

[10] G. Yang, H. Jain, A. Ganjoo, D. Zhao, Y. Xu, H. Zeng, and Guorong Chen, Opt. Express **16**, 10565 (2008).

[11] P. Němec, S. Zhang, V. Nazabal, K. Fedus, G. Boudebs, A. Moreac, M. Cathelinaud, and X.-H. Zhang, Opt. Express **18**, 22944 (2010).

[12] P. Hawlová, M. Olivier, F. Verger, V. Nazabal, P. Němec, Mater. Res. Bull. **48**, 3860 (2013).

[13] X. Su, R. Wang, B. L. Davies, and L. Wang, Appl. Phys. A **113**, 575 (2013).

[14] J. C. Phillips, J. Non-Cryst. Solids **34**, 153 (1979).

[15] H. He, and M. F. Thorpe, Phys. Rev. Lett. **54**, 2107 (1985).

[16] V. Benekou, L. Strizik, T. Wagner, S. N. Yannopoulos, A. L. Greer, and Jiri Orava, J. Appl. Phys. **122**, 173101 (2017).

[17] R. A. Street, Phys. Rev. B **17**, 3984 (1978).

[18] A. R. Barik, Mukund Bapna, D. A. Drabold, and K. V. Adarsh, Sci. Rep. **4**, 3686 (2014).

[19] X. Zhang, and D. A. Drabold, Phys. Rev. Lett. **83**, 5042 (1999).

[20] R. Sharma, K. Prasai, D. A. Drabold, and K. V. Adarsh, AIP Adv. **5**, 077164 (2016).

[21] H. Ftitzsche, Solid State. Commun. **99**, 153 (1996).

[22] M. Tatsumisago, B. L. Halfpap, J. L. Green, S. M. Lindsay, and C. A. Angell, Phys. Rev. Lett. **64**, 1549 (1990).

[23] R. Boehmer, and C. A. Angell, Phys. Rev. B **45**, 10091 (1992).

[24] P. Khan, H. Jain, and K. V. Adarsh, Sci. Rep. **4**, 4029 (2014).

[25] R. A. Street, Solid State Commun. **24**, 363 (1977).

[26] B. Ziaja, N. Medvedev, V. Tkachenko, T. Maltezopoulos, and W. Wurth, Sci. Rep. **5**, 18068 (2015).

[27] H. Harutyunyan, A. B. F. Martinson, D. Rosenmann, L. K. Khorashad, L. V. Besteiro, A. O. Govorov, and G. P. Weiderrecht, Nat. Nanotech. **10**, 770 (2015).

[28] M. Fadel, and S. S. Fouad, J. Mater. Sci. **36**, 3667 (2001).

[29] E. A. Davis, J. Non-Cryst. Solids **4**, 107 (1970).

[30] A. Regmi, A. Ganjoo, D. Zhao, H. Jain, and I. Biaggion, Appl. Phys. Lett. **101**, 061911 (2012).

[31] P. Khan, P. Acharja, A. Joshy, A. Bhattacharya, D. Kumar, and K. V. Adarsh, J. Non-Cryst. Solids **426**, 72 (2015).

[32] P. Khan, T. Saxena, H. Jain, and K. V. Adarsh, Sci. Rep. **4**, 6573 (2014).

[33] P. Lucas, J. Phys. Condens. Matter **18**, 5629 (2006).

**Figure captions**

**Fig. 1.** (a) Ground state optical absorption spectra of a-Ge$_x$As$_{35-x}$Se$_{65}$ thin films. Contour plot of the IA spectra of ⟨r⟩ = 2.60 sample for (b) 400 and (c) 560 nm excitation at various delays of the probe beam.

**Fig. 2.** Configurational Coordinate diagram showing two recombination paths of the excitons: (I) directly to the ground state (G), and (II) to a metastable state (M) via the creation of VAP.

**Fig. 3**. IA spectra of a-Ge$_x$As$_{35}$Se$_{65}$ thin films for 400 excitation at probe delays of (a) 1 and (b) 30 ps. DAS spectra associated with (c) DAS1 ($\tau_1$) and DAS2 ($\tau_2$) for all samples.

**Fig. 4**. IA spectra of ⟨r⟩ = 2.60 sample for different fluences of 560 nm excitation. (b) Variation of amplitude of IA for all samples in (b) bandgap and (c) sub-bandgap regions.

**Fig. 5.** (a) Timescales for predominant excitation and relaxation processes in a-Ge$_x$As$_{35-x}$Se$_{65}$ thin films. Time evolution of central wavelength for (b) DAS1 and (c) DAS2. (d) Dependence of DAS2 decay constant ($\tau_2$) against the MCN variation.

**Fig. 1**

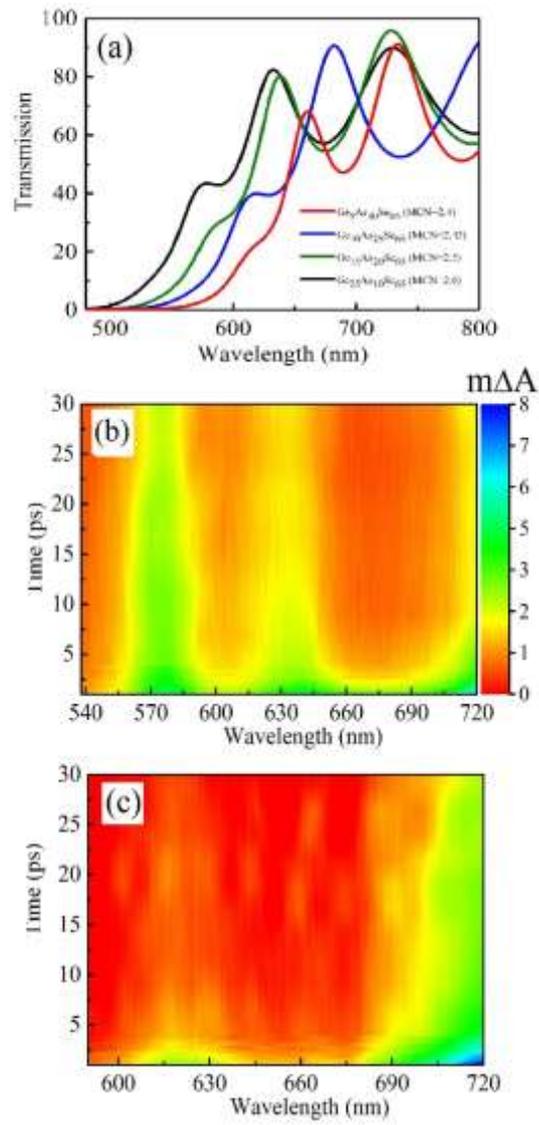

**Fig. 2**

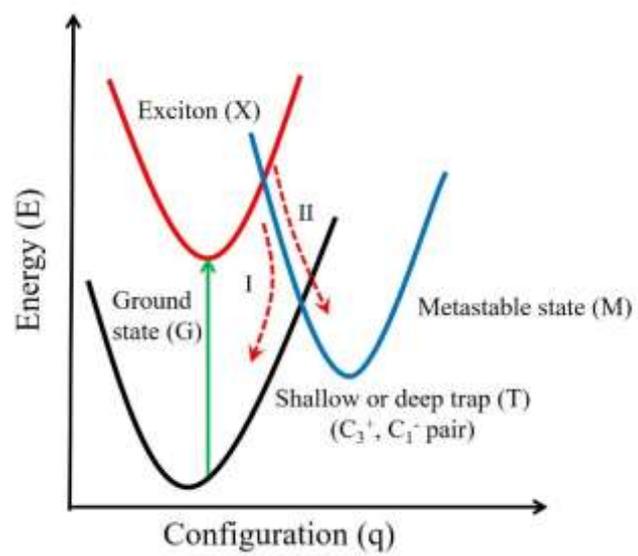

**Fig. 3**

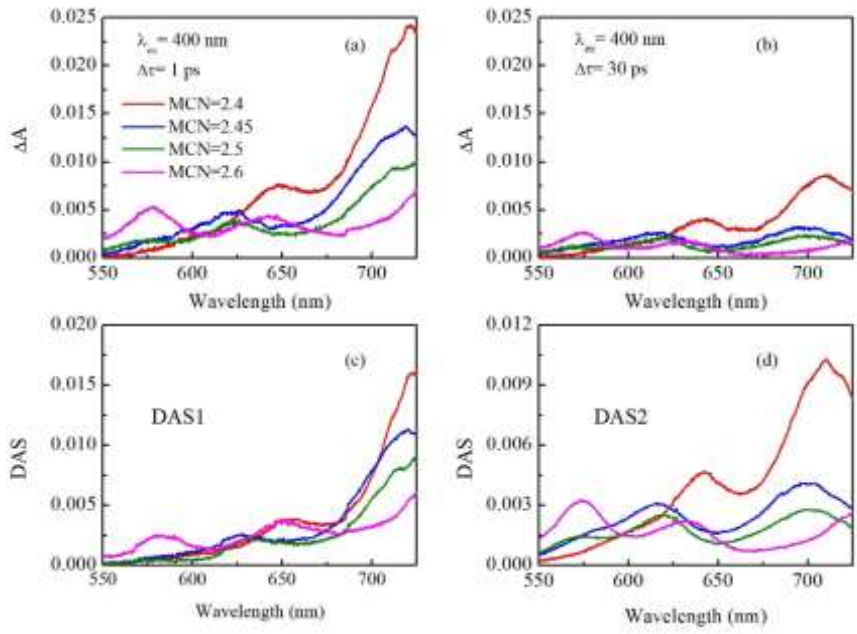

**Fig. 4**

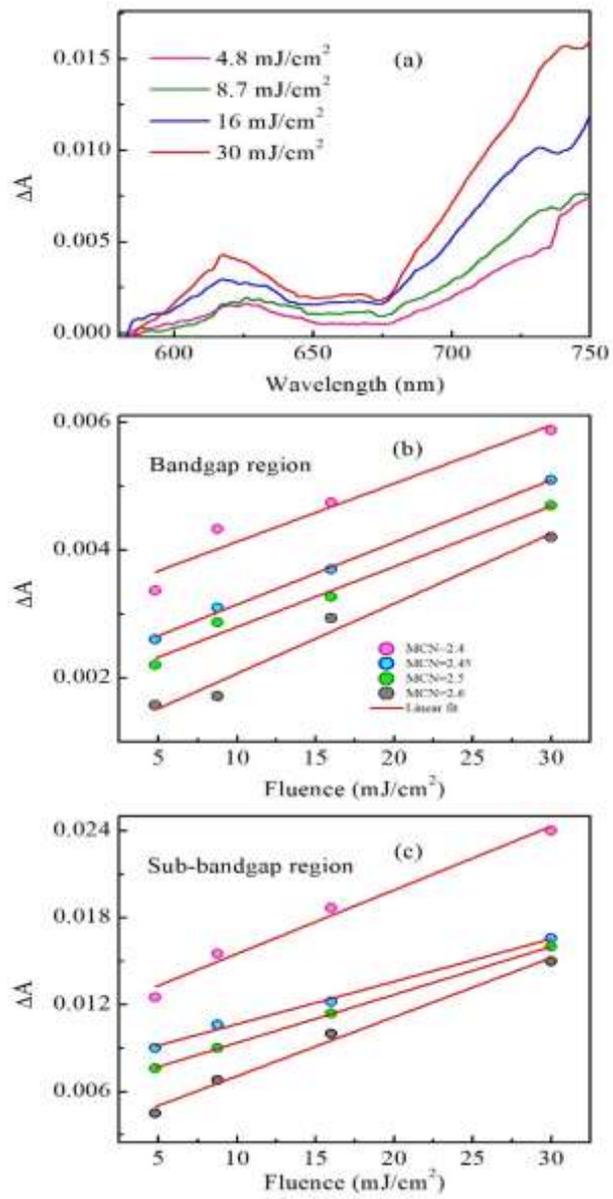

**Fig. 5**

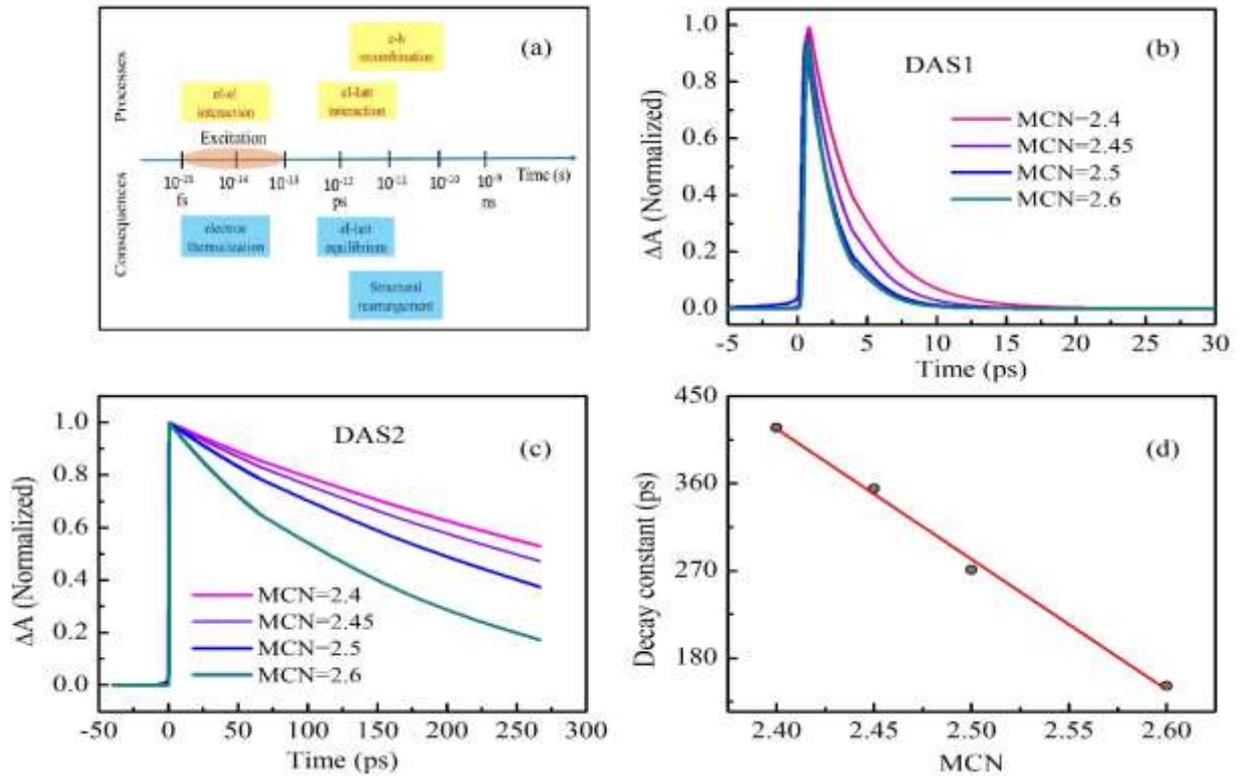